\documentclass[aps,preprint]{revtex4}
\usepackage{graphicx}

\begin{document}
\def\ba{\begin{eqnarray}}
\def\ea{\begin{eqnarray}}
\def\be{\begin{equation}}
\def\ee{\begin{equation}}
\def\la{\langle}
\def\ra{\rangle}
\def\h{\hskip 1cm}
\def\hh{\hskip 2cm}
\title{Exact solutions for universal holonomic quantum gates}

\author{V. Karimipour}
\email{vahid@sharif.edu}
 \affiliation{Department of Physics, Sharif University of
Technology, P.O.Box 11365-9161, Tehran, Iran}

\author{N. Majd}
\email{naymajd@mehr.sharif.edu}
 \affiliation{Department of Physics, Sharif University of
Technology, P.O.Box 11365-9161, Tehran, Iran}
\date{\today}

\begin{abstract}
We show how one can implement any local quantum gate on specific
qubits in an array of qubits by carrying adiabatically a
Hamiltonian around a closed loop. We find the exact form of the
loop and the Hamiltonian for implementing general one and two
qubits gates. Our method is analytical and is not based on
numerical search in the space of all loops.
\end{abstract}

\maketitle

\section{Introduction}\label{intro}
Since the suggestion of Zanardi and Rasetti in \cite{zr} on
holonomic quantum computation and its further development in
\cite{ ep, pzr, pz, pc, fuj}  concrete realization of holonomic
quantum gates has attracted a lot of attention. The attraction
comes mainly from the fact that this type of quantum computation
may be inherently stable against some local errors. The basic
idea of \cite{zr} is as follows. One takes an $n$ dimensional
Hamiltonian with a degenerate $k$ dimensional eigenspace
\begin{equation}\label{h(t)}
  H(t)|i(t)\ra = \epsilon (t) |i(t)\ra,
\end{equation}
where $\{|i(t)\ra, \ \ \  i=1\ \ \  k\} $ are the instantenous
degenerate eigenstates of the Hamiltonian. Carrying this
hamiltonian adiabatically around a loop in the parameter space,
induces a generalization of the Berry phase in the form of a
unitary operator in this subspace. This unitary operator is given
by
\begin{equation}\label{u}
  U= P e^{ - \oint_C A},
\end{equation}
where $P$ means path-ordered exponential, $A$ is the connection one form
\begin{equation}\label{i(t)}
  \la i(t) |A| j(t)\ra := \la i(t) | d |j(t)\ra, \h i,j=1 \ \ {\rm {to}} \ \
  k,
 \end{equation}
and $C$ is the loop around which the Hamiltonian is moved in the
parameter space. Here we are assuming that the basis vectors of
the Hilbert space have been arranged so that the first $k$ states
span the degenerate eigenspace of the Hamiltonian. \\ Of
particular interest to us is a scenario first proposed from a
geometrical point of view in \cite{thn}. The physical description
of the idea of \cite{thn} is as follows: To make the connection
one form time independent, one takes a family of iso-spectral
Hamiltonians on an $n$ dimensional Hilbert space defined as
follows:
\begin{equation}\label{x}
  H(t) = e^{tX}H_0e^{-tX},
\end{equation}
where $H_0=\epsilon \sum_{i=1}^{k} |i\ra\la i| \equiv \epsilon
P_0$, and $X$ is an anti-hermitian operator acting on the Hilbert
space. Thus $P_0$ is the projection operator for the $k-$
dimensional eigenspace, and the operator $X$ determines the
adjoint action and hence the curve in the parameter space. Note
that the curve has been parameterized by the real parameter $t\in
[0,1]$ and if one requires that the curve be closed in the
parameter space, one should have
\begin{equation}\label{closedcurve}
  P_0 = e^{X}P_0e^{-X}.
\end{equation}
It is always possible to choose the basis of the Hilbert space so
that
\begin{equation}\label{p0}
  P_0 = \left(\begin{array}{cc} I_{k\times k} & 0 \\ 0 & 0 \end{array}\right).
\end{equation}
For the Hamiltonian (\ref{x}) we have:
\begin{equation}\label{state}
  |i(t)\rangle = e^{tX} |i\ra,
\end{equation}
from which we obtain the time independent connection
\begin{equation}\label{ax}
  \la i(t) |A|j(t) \ra = \la i |X|j\ra, \h i, j = 1 \ \ \cdots \ \ \  k.
\end{equation}
Note that this equation by no means implies the $A=X$. It only
means that the restriction of the operator $X$ to the
degenerate subspace is equal to $A$.\\
Since $X$ is constant, calculation of the path integral is quite simple and gives
\begin{equation}\label{u2}
  U = P e^{-\oint_C A} = e^{-A}.
\end{equation}
Note that the operator $X$ determines the curve in the parameter
space completely. However only its projection to the degenerate
subspace is determined by the operator $U=e^{-A}$ which we want to
implement $\varpi$ holonomically through equation (\ref{ax}), and
one is free to choose the rest of this operator to satisfy the
constraint (\ref{closedcurve}) which means that the curve should
be a closed one. As an explicit example \cite{thn}, if one wants
to implement a one qubit gate $u=e^{-A} \in U(2)$  one should take
$k=2$ and can take $n=3$ and $X$ as
\begin{equation}\label{example}
  X = \left(\begin{array}{cc}
    A & w \\
    -w^{\dagger} & 0 \
  \end{array}\right),
\end{equation}
where $w$ is a two dimensional vector which should be determined
by satisfying the constraint (\ref{closedcurve}).  In \cite{thn}
the dimension $n$ is always taken to be $k+1$ and solutions for
some one quibt gates (e.g. the Hadamard gate) for which $k=2$ and
two qubit gates (e.g. the CNOT gate and the 2 dimensional fourier
transform) for which $k=4$ are obtained by extensive numerical
search in the space of two and four dimensional vectors
respectively.\\
Our aim in this paper is two-fold. First we provide an exact
method (free of numerical calculations) for obtaining solutions
for arbitrary unitary gates acting on an arbitrary number of
qubits. We obtain in a much simpler method the solutions already
obtaind in \cite{thn}. Second we address the problem of
scalability of this construction. That is, we modify the method
of \cite{thn} so that we can holonomically implement any local
quantum gate on an array of qubits by appropriate choices of
curves in the parameter space. This will give us full power for
doing holonomic
computation on a quantum computer in a scalable way.\\
The structure of the paper is as follows: In section
\ref{section2} we present our exact solution for the case when
$n=k+1$. Since this is the case already studied by the authors in
\cite{thn}, who have supplied various examples by searching in
the space of matrices numerically, we suffice to give only proofs
for general gates and refrain from giving examples. In section
\ref{scale} we generalize our method and show how to
holonomically implement any local quantum gate on an array of
qubits by appropriate choices of curves in the parameter
space. \\ We end the paper by a short discussion.\\

\section{Analytical solution}\label{section2}
Let us take an arbitrary anti-hermitian operator $X\in u(n)$.
This determines the adjoint orbit via equation (\ref{x}). Due to
equation (\ref{ax}) and the fact that $n=k+1$, the general form
of $X$ is as follows:
\begin{equation}\label{formofx}
  X=\left(\begin{array}{cc} A & w \\ -w^{\dagger} & i\ s
  \end{array}\right),
\end{equation}
where $s$ is a real number and $i=\sqrt{-1}$.
 Note that the upper $k$ by $k$ diagonal block of $X$ is
dictated by the choice of unitary operator $U=e^{-A}$ that we
want to represent holonomically. We  have choices only for the
vector $w$ and the number $s$. We can use this
freedom to solve the constraint (\ref{closedcurve}) which requires that the loop be closed. \\
In \cite{thn} the parameter $s$ has been taken equal to zero.
However we will see that a better choice for this parameter
exists which will simplify the final results considerably. Our
task is now to exponentiate the operator $X$.\\ Exponentiation of
a specific anti-hermitian operator is a straightforward (albeit
highly tedius) task which can be done by diagonalization of the
operator (since we have to solve cubic equations with general
coefficients). Since we want to do this for a completely general
operator of the form (\ref{formofx}), for which we have no a
priori information about the matrix $A$ and the vector $w$, we
make the simplifying restriction that $w$ is proportional to an
eigenvector of $A$, that is
\begin{equation}\label{w}
  Aw = i\lambda w, \h w^{\dagger}w = \alpha^2,
\end{equation}
where $\lambda $ is a real parameter.\\
With this assumption we can use a recursive method for
calculation of $e^{tX}$ instead of numerical methods. \\
A simple
calculation for the first few powers shows that the general form
of $X^n$ is
\begin{equation}\label{xn}
  X^n=\left(\begin{array}{cc} A^n - b_n ww^{\dagger} & c_n w \\ -c^*_n w^{\dagger} & d_n
  \end{array}\right),
\end{equation}
where $a(n), b(n)$ and $ c(n)$ are coefficients to be determined.  These coefficients satisfy the following
initial conditions:
\begin{eqnarray}\label{initial}
  b_0 &=& 0 \ \ \  b_1 = 0,  \cr
  c_0 &=& 0 \ \ \  c_1 = 1,  \cr
  d_0 &=& 1 \ \ \  d_1 = i\ s .
\end{eqnarray}
To obtain these coefficients we use
\begin{eqnarray}\label{xn+1}
  X^{n+1}&=&\left(\begin{array}{cc} A & w \\ -w^{\dagger} & i\ s \end{array}\right)\left(\begin{array}{cc}
   A^n - b_n w w^{\dagger} & c_n w \\ -c^*_n w^{\dagger} & d_n
  \end{array}\right)\cr
&=&\left(\begin{array}{cc} A^{n+1} - b_{n+1} ww^{\dagger} &
c_{n+1} w
\\ -c^*_{n+1} w^{\dagger} & d_{n+1}
\end{array}\right),
\end{eqnarray}
from which we obtain the following recursive system of equations
\begin{eqnarray}\label{anbncn}
  b_{n+1} &=& i\lambda b_n + c_n^*,\cr
  c_{n+1} &=& i\lambda c_n + d_n, \cr
  d_{n+1} &=& i s d_n -\alpha^2 c_n.
\end{eqnarray}
To solve this system we define the following generating functions
\begin{eqnarray}\label{BCD}
  B(t)&:=& \sum_{n=0}^{\infty} \frac{b_n}{n!}t^n,\cr
  C(t)&:=& \sum_{n=0}^{\infty} \frac{c_n}{n!}t^n,\cr
  D(t)&:=& \sum_{n=0}^{\infty} \frac{d_n}{n!}t^n.
\end{eqnarray}
Once we find these generating functions we have the exponential as
\begin{equation}\label{etx}
  e^{tX} = \left(\begin{array}{cc} e^{tA}-B(t)ww^{\dagger}& C(t)w \\ - C^*(t)w^{\dagger}& D(t)\end{array}\right).
\end{equation}
In terms of generating functions the recursion relations read as
follows:
\begin{eqnarray}\label{B'C'D'}
  B'(t) &=& i\lambda B(t) + C^*(t),\cr
  C'(t) &=& i \lambda C(t) + D(t), \cr
  D'(t) &=& i s D(t) - \alpha^2 C(t).
\end{eqnarray}
The second and the third equations can be combined to give
\begin{equation}\label{C''}
  C''(t) = i(\lambda + s) C'(t) + (\lambda s - \alpha^2) C(t),
\end{equation}
the solution of which is
\begin{equation}\label{C(t)}
  C(t) = \gamma_1 e^{q_1 t} + \gamma_2 e^{q_2 t},
\end{equation}
where
\begin{equation}\label{q12}
  q_{1,2} =  \frac{i}{2}(\lambda+s \pm \sqrt{(\lambda-s)^2 + 4 \alpha^2})=: i(\frac{\lambda +s}{2} \pm \nu),
\end{equation}
where we have defined $\nu
:=\frac{1}{2}\sqrt{(\lambda-s)^2+4\alpha^2} $ and $\gamma_{1,2}$ are constants to be determined from initial conditions. \\
 In view of the initial conditions
\begin{equation}\label{c0}
  c_0 \equiv C(0) = 0  \ \ \ \ \ {\rm{and}} \ \ \ \ \  c_1 \equiv C'(0) =
  1,
\end{equation}
we find the final form of the generating function
\begin{equation}\label{C(t)final}
  C(t) = \frac{e^{q_1 t} - e^{q_2 t}}{q_1 - q_2} = e^{i\frac{(\lambda+s) t}{2}} \frac{\sin \nu t}{\nu}.
\end{equation}
From this solution we find that a simplifying choice for the
parameter $s$ is to take $ s = - \lambda $ for which we will
have:
\begin{equation}\label{C(t)simplified}
  C(t) = \frac{\sin \nu t}{\nu},
\end{equation}
where now
\begin{equation}\label{nusimplified}
\nu = \sqrt{\lambda^2 + \alpha^2}
\end{equation}.
Equation (\ref{B'C'D'}) will then give
\begin{equation}
  D(t) = C'(t) - i\lambda C(t) = \cos \nu t -
  i\frac{\lambda}{\nu}\sin \nu t.
\end{equation}
Finally we solve the first equation of (\ref{B'C'D'}) by rewriting
it as
\begin{equation}
  \frac{d}{dt} \left(e^{-i\lambda t} B(t)\right) = e^{-i\lambda t} C^*(t),
\end{equation}
the solution of which is
\begin{eqnarray}
  B(t) &=& \int_0^{t} e^{i\lambda(t-\tau)} C^*(\tau) d\tau\cr
  &=& \frac{1}{(\nu^2-\lambda^2)}\left(e^{i\lambda t} -\cos \nu t - i \frac{\lambda}{\nu}\sin \nu t \right),
\end{eqnarray}
where we have used the initial condition $ b_0 \equiv B(0) = 0 $. \\
What we now require is that the constraint (\ref{closedcurve}) be
satisfied. This constraint is satisfied when the operator $e^X$
becomes block-diagonal, since in this case $[e^X, P_0]=0$. Due to
equation (\ref{etx}) this requires that we set $ C(1) = 0 $ which
in view of (\ref {C(t)final})
 demands that we fix  $\nu $ as $ \nu =
\pi n $ or in view of equation (\ref{nusimplified}), $ \alpha^2 = (n\pi)^2 - {\lambda^2}$, where $n$ is an integer.\\
Thus the recepie is as follows: for any given operator
$U=e^{-A}$, we choose an eigenvector $v$ of $A$, we form the
matrix $X$ as in (\ref{formofx}) with $w=\alpha v$ where $\alpha
$ is given as $\alpha^2 = (n\pi)^2 - \lambda^2 $. Thus for any
operator in a $k$ dimensional space there are a number of $k$
closed curves, each corresponding to one of the eigenvectors of
$A$. In view of equation (\ref{C(t)final}), for each fixed $n$
the Hamiltonian $H(t)$ turns back to its original form $H_0$
exactly $n$ times during the period from
$t=0$ to $ t=1$.  The optimal choice for $n$ is thus $n=1$. \\

\section{Generalization and physical realization}\label{scale}
Let us summarize our findings up to now. If we want to implement
a unitary operator (a quantum gate) on a $k$ dimensional Hilbert
space, we take a system whose Hilbert space is  $k+1$
dimensional. On this larger Hilbert space, we take a Hamiltonian
$H_0=P_0$, where $P_0$ is the projector operator on the smaller
$k$ dimensional system and then adiabatically transform this
initial Hamiltonian by the adjoint action $H(t)= e^{tX}P_0
e^{-tX}$, where $X$ is a vector in the lie algebra $\in u(k+1)$.
For any given unitary operator $U\in U(k)$, we have a suitable
vector $X$, such that the holonomy operator which is
adiabatically induced on the degenerate $k$ dimensional Hilbert
space coincides with that operator, that is $ e^{-\oint_C A}= U$,
where $\la i(t)|A|j(t)\ra = \la i(t)|d|j(t)\ra = \la i|X|j\ra , \
\ \ \ \  1\leq i, j \leq k $.\\
Actually for any given operator $U$, we have $k$ generally
different solutions $X$, since we have $k$ different choices for
the vector $w$ to insert in (\ref{formofx}), each proportional to
a different eigenvector of the operator $A$. This is an
important point which we will come to later in this section.
In fact we will see that it is best to combine the separate solutions into one single solution. \\
The reason is that when it comes to physical realization of the
technique in section (\ref{section2}), the problem of dimension
becomes an annoying obstacle, since if we want to implement a
single qubit gate, we require a three dimensional Hilbert space
and if we want to implement a two-qubit gate, we need a five
dimensional Hilbert space and so on. Generally we do not have
such a control over the dimensions of Hilbert spaces in most
experimental scenarios for quantum computation. In most of these
scenarios we have a number $n$ of two dimensional systems (e.g.
spins, ions , etc.) which play the role of qubits which when put
together make a Hilbert space of dimension $2^n$. When one
enlarges this system by adding a quibt (an ancilla qubit) the
dimension of the Hilbert space doubles to $2^{n+1}$ instead of
getting increased by one. \\
The most ideal situation is to have a main array of qubits
numbered from $1$ to $n$ together with an ancilla qubit which we
number as $n+1$. This ancilla qubit is then used for implementing
holonomically a set of universal gates (one qubit and two qubit
gates) on any qubit or pair of qubits in the main array
of qubits.\\
In order to realize this ideal situation we modify our method as
follows. We let the dimension of the larger Hilbert space to be
twice the dimension of the degenerate subspace. For simplicity of
presentation we first consider the example where we have a single
qubit as our main system and a single ancilla qubit. Instead of
(\ref{formofx}) we now take $X$ as follows:
\begin{equation}\label{formofx2}
  X = \left(\begin{array}{cc}
    A& \left(w_1, w_2\right) \\
    -\left(\begin{array}{c}
      w_1^{\dagger} \\
      w_2^{\dagger} \\
    \end{array}\right) & \left(\begin{array}{cc} -i\lambda_1 & 0 \\ 0 & -i\lambda_2\end{array}\right) \
  \end{array}\right),
\end{equation}
where $w_1=\alpha_1 v_1 $ and $ w_2=\alpha_2 v_2$ are
proportional to the two eigenvectors of the operator $A$
corresponding to eigenvalues $\lambda_1$ and $\lambda_2$
respectively. In this way we incorporate the two separate
solutions into one
solution for the sake of scalability of holonomic computation.\\
We can easily repeat all the calculations in section
(\ref{section2}). For example we will have
\begin{equation}\label{xn2}
  X^n = \left(\begin{array}{cc}
    A^n - b_{1,n}w_1  w_1^{\dagger} - b_{2,n} w_2 w_2^{\dagger} & \left(c_{1,n}w_1, c_{2,n}w_2\right) \\
    -\left(\begin{array}{c}
      c^*_{1,n} w_1^{\dagger} \\
      c^*_{2,n} w_2^{\dagger} \\
    \end{array}\right) &  \left(\begin{array}{cc} d_{1,n}& 0 \\ 0 & d_{2,n}\end{array}\right)\
  \end{array}\right),
\end{equation}
where the functions $b_{k,n}, c_{k,n}$ and $ d_{k,n}$ for $k=1,2$
are found to satisfy the following recursion relations:
\begin{eqnarray}\label{bncndn2}
  b_{k,n+1} &=& i\lambda_k b_{k,n} + c^*_{k,n},\cr
  c_{k,n+1} &=& i\lambda_k c_{k,n} + d_{k,n},\cr
  d_{k,n+1} &=& -i\lambda_k d_{k,n} - \alpha_k^2c^*_{k,n}.
\end{eqnarray}
This shows that the recursion relations for various indices $k$
are decoupled and thus the solutions for the generating functions
found in section (\ref{section2}) can be directly carried over to
here. For example we have for each $k$, that
\begin{equation}\label{Ck(t)}
  C_k(t) : = \frac{\sin \nu_k t}{\nu_k},
\end{equation}
where  $ \nu_k = \sqrt{\lambda^2+\alpha_k^2} $. This tells us
that the matrix $e^{X}$ will be block diagonal if we choose
$\nu_k = n_k \pi $ or
optimally if $\nu_k = \pi$. \\
A more concise and instructive way of doing the above calculation
is to rewrite $(w_1, w_2, \cdots w_n)$ as
\begin{equation}
(w_1, w_2, \cdots w_n) = (\alpha_1 v_1, \alpha_2 v_2, \cdots
\alpha_n v_n) =\Omega D,
\end{equation}
where $\Omega$ is the unitary matrix which diagonalizes $A$ and
$D = {\rm {diagonal}}(\alpha_1, \alpha_2, \cdots \alpha_n)$. Then
we can write $X$ in the form
\begin{equation}\label{xconcise}
  X = \left(\begin{array}{cc}
    A& \Omega D \\
    -D\Omega^{\dagger} & -i\Lambda\
  \end{array}\right) =  \left(\begin{array}{cc}
    \Omega & 0 \\
    0 & I \
  \end{array}\right)\left(\begin{array}{cc}
    i\Lambda & D \\
    -D & -i\Lambda \
  \end{array}\right)\left(\begin{array}{cc}
    \Omega^{\dagger} & 0 \\
    0 & I \
  \end{array}\right).
\end{equation}
In this form the matrix $X$ is easily exponentiated to give
\begin{equation}\label{etxconcise}
  e^{tX} = \left(\begin{array}{cc}
    \Omega & 0 \\
    0 & I \
  \end{array}\right)e^{tX_0 }\left(\begin{array}{cc}
    \Omega^{\dagger} & 0 \\
    0 & I \
  \end{array}\right),
\end{equation}
where the matrix $X_0$ has been defined as
\begin{equation}\label{x0}
 X_0:=\left(\begin{array}{cc}
    i\Lambda & D \\
    -D & -i\Lambda
  \end{array}\right).
\end{equation}
It is clear that
\begin{eqnarray}\label{x0x02}
  X_0 &=& i\left(\sigma_z \otimes \Lambda + \sigma_y \otimes D\right),\cr
  X_0^2 &=& -I\otimes (\Lambda^2 + D^2).
\end{eqnarray}
Note that both $\Lambda $ and $D$ are diagonal matrices. This
then leads to the following result
\begin{equation}\label{etx0concise}
  e^{X_0t} = I \otimes \cos \nu t + i(\sigma_z\otimes \Lambda + \sigma_y\otimes D) (I\otimes \frac{\sin \nu
  t}{\nu}),
\end{equation}
where $\nu:= \sqrt{\Lambda^2 + D^2}$ is a diagonal matrix which
generalizes and replaces the simple factors $\nu =
\sqrt{\lambda^2 + \alpha^2}$ of section (\ref{section2}). The
condition (\ref{closedcurve}) is now satisfied if we choose
\begin{equation}\label{nuk}
  \nu_k = \sqrt{\lambda_k^2 + \alpha_k^2} = n_k \pi \ \ \ \ \  \forall
  k.
\end{equation}
If we take all these factors equal to $\pi$ which is the optimal
choice for each factor, the
matrix $\nu$ will equal $\pi I$. This brings about further simplifications as we will see.\\
From (\ref{etxconcise}) and (\ref{etx0concise}), we find that
\begin{equation}\label{etxsimplifiedfinal}
  e^{tX} =\left(
\begin{array}{cc}
  \cos \pi t + \frac{A}{\pi}\sin \pi t&  \frac{\Omega D}{\pi} \sin \pi t \\
   -\frac{D\Omega^{\dagger}}{\pi} \sin \pi t &\cos \pi t - i\frac{\Lambda}{\pi}\sin \pi t
\end{array}
  \right).
\end{equation}
Now that we have obtained the general form of the operator $X$,
for any unitary operator, let us see what is its form for a one
qubit gate $u:=e^{a}$ acting on one of the qubits, say qubit $k$,
in the main array.   Then we will have
\begin{eqnarray}\label{ak}
  A &=&  I^{\otimes k-1}\otimes a\otimes I^{\otimes^{n-k}}, \ \ \ \ \
  \Omega = I^{\otimes k-1}\otimes \varpi \otimes I^{\otimes^{n-k}}, \cr
  \Lambda &=& I^{\otimes k-1}\otimes \lambda\otimes I^{\otimes^{n-k}}, \ \ \ \ \
  D = I^{\otimes k-1}\otimes d\otimes I^{\otimes^{n-k}},
\end{eqnarray}
where $\varpi$ is the matrix which diagonalizes $a$, $\lambda$ is
the diagonalization of $a$, and $d$ is the diagonal matrix $d=diag
(\alpha_1, \alpha_2)$, where $\pi =\sqrt{\lambda_k^2 +
\alpha_k^2}$. Thus we see that for a local one qubit gate
$u_k=I^{\otimes {k-1}}\otimes u \otimes I^{\otimes {n-k}}$ which
acts trivially on all other qubits except the $k-$th qubit, our
matrix $X$ will also be a local matrix
\begin{equation} X_k:=I^{\otimes {k-1}}\otimes x \otimes I^{\otimes
{n-k}},\end{equation} is a matrix acting on the $k-th$ qubit and
the ancilla and $x$ is a four by four matrix of the form
(\ref{xconcise}), that is $x=\left(\begin{array}{cc}
  a & \varpi d \\
  -d\varpi^{\dagger} & -i\lambda
\end{array}\right)$. \\
By the same analysis, we find that a local two qubit gate can
also be implemented holonomically, by a local choice of $X$
acting only on the two qubits and the ancilla. The general form
of the matrix $X$ is always as given in (\ref {xconcise}).
\\
For $H_0$ we can assume the following form $H_0 =
\frac{1}{2}(I+\sigma_z)\otimes I$, where the first factor acts on
the ancilla and the second factor acts on all the other qubits.
Thus the Hamiltonian $H_0$ describes an array of qubits in which
only the ancilla qubit is in a magnetic field in the $z$
direction. The eigenstates of $H_0$ are $|0\ra\otimes |e_k\ra, k=1
\ \ \ to \ \ \  2^n$ with energy $E=1$ and $|1\ra \otimes |e_k\ra,
k=1 \ \ \ to \ \ \  2^n$ with energy $E=0$, where $|e_k\ra$ are
the basis vectors of the main array of qubits. Thus the
Hamiltonian is the projection operator on the main array.
According to (\ref{state}), one now moves arround the orthonormal
frame $\{|0\ra \otimes |e_k\ra\}$, by the action of the operator
$e^{tX}$ which means that the Hermitian operator $iX$ acts like a
Hamiltonian in the fixed frame. After one cycle of revolution from
$t=0$ to $t=1$, the desired operator corresponding to the choice
of $X$ has been implemented holonomically on the main
array.\\
For a one qubit gate, the operator $x$ can be written in terms of
Pauli matrices in the form
\begin{equation}\label{xsigma}
  x = \frac{1}{2}(I+\sigma_z)\otimes a - \frac{i}{2}(I-\sigma_z)\otimes\lambda
  + \sigma_+\otimes \varpi d - \sigma_- \otimes d\varpi^{\dagger}.
\end{equation}
Here the first factor of the tensor product acts on the ancilla
and the second factor acts on a local qubit. As a concrete example
if we take $u=\sigma_z$, for which $a=\frac{i\pi}{2}(I-\sigma_z)$
and $\varpi = I$, and follow the above construction, after
straightforward calculations we find the following form of the
operator $x$:
\begin{equation}\label{xconcrete}
  x = \frac{i\pi}{2} \left[(\sigma_z -\sigma_y)\otimes I + (\sigma_y-I)\otimes
  \sigma_z\right].
\end{equation}

\section{Discussion}
We have found exact solutions for the holonomic implementation of
local quantum gates on an arbitrary array of qubits plus an
ancilla. A family of isospectral Hamiltonians all with the same
spectrum of the projection operator on the main array (without
the ancilla) is constructed by the choice of a suitable curve in
the form of $H(t)=e^{tX}H_0e^{-tX}$, defined by an operator in
the full Hilbert space of the array and the ancilla.  For every
single qubit gate acting on the $k-$th qubit, there is a local
operator $X_k$ acting nontrivially only on the ancilla and the
qubit in question, which implements that gate on the $k-$th qubit
holonomically. This is also the case for two qubit gates.\\
Thus by imbedding the appropriate operators for a suitable set of
one and two qubit gates which comprize a universal set of quantum
gates, in various places in the tensor product of the array, one
can enact holonomically any set of gates on any subset of qubits. \\
This of course leads us to the actual physical implementation of
such a method in a concrete physical realization of quantum
computers. We leave this problem for future research.\\ \\
\textbf{Acknowledgement} We hereby thank A. Sadrolashrafi and A.
T. Rezakhani for valuable discussions. After submitting this paper
we were informed by the authors of \cite{thn} that they have
announced a similar solution (without proof) in their revised
paper.

\end{document}